# Superexcitability in Chiral Oscillators with Spatial Degrees of Freedom

Alessandro Scirè[1*]

1: University of Pavia, Department of Electrical, Computer and Biomedical Engineering
via A. Ferrata 5 – 27100 – Pavia (Italy).
*: Correspondence: alessandro.scire@unipv.it;

**Abstract**

This work presents a numerical investigation of interacting chiral oscillators (COs), characterized by an intrinsic rotational handedness. The coupling of positional and orientational degrees of freedom drives a mutual influence between synchronization and spatial dynamics. The control parameter, Δ (anisotropy or detuning), represents the frequency difference between two COs (for the two-body case) or between two families of COs (for the many-body case) with opposite handedness. For two COs with opposite handedness, interaction forces generate a Helical Excitable Dipole (HED), where excitability is a spatiotemporal act characterized by a helical trajectory. For many COs moving in 2Ds, the model displays a locking-to-unlocking transition of the Kuramoto kind driven by Δ, and, within the locking region, many "superexcitable" states sustained by a global saddle-node bifurcation, detected by the collective period behavior versus Δ. These coherent, topological states are characterized by dissipationless phase-momentum locking, which I quantified using appropriate global metrics, including the Kuramoto order parameter and the novel parameter S capable of detecting the phase-momentum locking. These states exhibit a wide variety of topologically protected vortex complexes, whose complexity increases with system size. The model provides a unified framework for diverse biophysical applications—from molecular ratchets to cardiac looping—identifying the symmetry breaking of motile excitable units as a fundamental process for large-scale, robust collective transport.

## 1. Introduction

A fundamental question in the study of complex systems is: how can local, sub-threshold interactions give rise to long-range coherent collective states? Furthermore, how do locally coupled excitable units self-organize when their internal state (phase) is coupled to their physical position? In classical models space often acts as a passive background or a fixed lattice. However, in many biophysical contexts—such as molecular motors or swimming microorganisms—the "firing" of a unit is not merely a temporal event but a mechanical act that involves displacement. Despite great progress in understanding excitability, its spatial self-organization still reserves significant avenues for research, particularly regarding how locally interacting, sub-threshold excitable elements can unite to exhibit spatially distributed, coherent behavior.

Excitability is defined as the ability of a system to undergo a large, non-linear excursion in response to a perturbation that exceeds a specific threshold, while returning quickly to a quiescent state when faced with sub-threshold stimuli. This phenomenon is ubiquitous across diverse scientific domains, appearing in the non-linear dynamics of semiconductor lasers [1], the chemical oscillations of Belousov-Zhabotinsky reactions [2], in the large-scale spreading of wildfires [3] or social information [4]. However, it is in the study of living matter that these principles find their most intricate expression, ranging from the rhythmic beating of the heart [5] to neuronal avalanches [6], from the collective migration of epithelial tissues [7] to the metachronal coordination of cilia [8].

Over the last decades, the study of excitability has evolved from classical FitzHugh-Nagumo [9] or Hodgkin-Huxley [10] models toward the understanding of increasingly complex collective states. Current research focuses not only on the response of single elements but on how thermal noise [11], parameter heterogeneity [12], and connection topology [13] can induce global coherence phenomena, such as stochastic resonance [14] or phase synchronization [15]. Recent findings in active matter and biological tissues have highlighted that the excitable response is not merely a temporal event, but a dynamic phase transition capable of generating self-organized propagation waves and persistent spatial patterns, where energy is stored and released to perform coordinated mechanical work [16]. In this field, researchers have quantified how intrinsic cellular chirality translates into large-scale collective motion, revealing a fundamental mechanism for biological morphogenesis [17]. Specifically, theoretical studies have explored singular density correlations in chiral active fluids in three dimensions, alongside investigations into the collective motion of self-trapping chiral active particles [18]. However, considerable scope remains for future research into how the principles of cellular chirality and collective excitability are precisely integrated during complex morphogenetic processes.

In this work, I propose a framework where spatial and phase degrees of freedom are entangled through an intrinsic geometry. I introduce the Helical Excitable Dipole (HED), formed by a pair of counter-rotating Chiral Oscillators (COs) detuned by a form of circular anisotropy, and where the internal phase dynamics are coupled to the spatial coordinates, making each excitation cycle a "helical pump" with a quantitative spatial consequence. Building upon this dipolar unit, I define "superexcitability" as an emergent phase that reflects a collective manifestation of excitability in the many-body case. Unlike classical collective excitability, which often concerns the propagation of pulses in locally coupled excitable media, superexcitability is governed by a global saddle-node bifurcation within the parameter regime where local excitability remains sub-critical. The global saddle-node bifurcation enables the emergence of robust, topologically protected, supercritical vortex complexes, the superexcitable states. The prefix "super" highlights the topological, coherent, and dissipationless nature of these patterns, which also exhibit phase-momentum locking, in analogy with the robust transport found in topological insulators or superconductors. The phase diagram reveals that this transition is governed by a specific threshold condition involving the product of anisotropy and system size.

The manuscript is organized as follows. In Section 2, I introduce the model for two Chiral Oscillators (based on the well known Adler equation) and define the dynamics of the Helical Excitable Dipole (HED) deriving the geometric properties of its "helical pump" mechanism. Section 3 explores the many-body system using numerical results, where I characterize the emergence of the superexcitable phase, analyze the topological vortex-antivortex complexes and the emergence of ballistic transport through phase-momentum locked loops, and present the resulting phase diagram in the parameter space, discussing the threshold condition for superexcitability. Finally, Section 4 discusses the theoretical aspects of superexcitability and its biophysical applications—ranging from molecular motors to cardiac morphogenesis—and provides concluding remarks. Appendix A details the stability analysis of the HED, and Appendix B examines the interdipolar forces.

## 2. The Helical Excitable Dipole

As a starting point I address local excitability by employing the Adler equation for two coupled oscillators, before any spatial degrees of freedom are introduced. The Adler model (published by Robert Adler in 1946 to explain injection locking in electronic LC oscillators) provides the simplest description of a phase-excitable system, where an external perturbation can trigger a full $2\pi$ rotation—an elementary 'firing' event. It is used to model phenomena in fields as diverse as neuroscience (firing patterns of neurons [20]), biology (circadian rhythms [21], coupling of beating cilia and flagella [22]), superconductivity (Josephson junctions [23]), and photonics (laser synchronization [24]).

I consider two phase oscillators, a counterclockwise (CCW) $\varphi_+$ and a clockwise (CW) $\varphi_-$, both living in $S^1$ and sinusoidally coupled. With unit coupling and natural frequencies detuned as $\omega_+ = \Delta$ and $\omega_- = -\Delta$, with $\Delta$ non-negative without losing generality. The equations of motion are:

$$\dot{\varphi}_+ = \Delta + \sin(\varphi_- - \varphi_+), \quad (1)$$

$$\dot{\varphi}_- = -\Delta + \sin(\varphi_+ - \varphi_-). \quad (2)$$

The Adler equation for the relative phase $\phi = \phi_+ - \phi_-$ is

$$\dot{\phi} = 2\Delta - 2\sin(\phi). \quad (3)$$

If the system synchronizes ($\dot{\phi} = 0$), the stationary phase is determined by the condition

$$\sin(\phi) = \Delta. \quad (4)$$

When the driving force $\Delta$ is below a critical value ($\Delta < 1$), the system settles into a stable resting state (synchronization). However, if the drive exceeds the threshold ($\Delta > 1$), the system loses its stable fixed point via Saddle-Node (SN) bifurcation and enters a limit cycle—it "fires" a continuous train of phase slips. The threshold at ($\Delta = 1$) is the critical point where the system switches from having a stable fixed point to a limit cycle.

Despite being a single-phase problem, the Adler equation (3) is not actually describing a single oscillator, but the collective behavior of a pair of oscillators with opposing chiralities, where I define chirality as the intrinsic sense of rotation (CW or CCW) of the phase in the internal parameter space $S^1$. It is also worth noticing that Eqs. (1)-(2) represent the Kuramoto [25] model with N = 2, while they also describe an excitable "dipole" encapsulated by the single variable $\phi = \phi_+ - \phi_-$.

To extend the theory toward spatialized structures, I introduce the Helical Excitable Dipole (HED) consisting of chiral oscillators of opposite chirality whose dynamics are governed by the coupled evolution of a relative phase $\phi \in S^1$ and a relative spatial coordinate $x \in \mathbb{R}$, on a cylindrical manifold $\mathbb{R} \times S^1$. The equations of motion are defined as:

$$\begin{cases} \dot{x} = W'(x)\cos(\phi) & (5) \\ \dot{\phi} = \Delta - W(x)\sin(\phi). & (6) \end{cases}$$

Here, $W(x)$ is a short-range interaction kernel that links the internal phase dynamics and the spatial configuration; $\Delta$ is the intrinsic frequency detuning. A detailed stability analysis of Eqs. (5)-(6) is reported in Appendix A, in the following I deal with the excitability properties.

The interaction kernel is chosen (see discussion at the end of subsection 3.1) as $W(x) = -\exp(-x^2)$, so that the characteristic interaction length L = 1. The dynamics of Eqs.(5)-(6) are governed by a tilted washboard potential:

$$V(x, \phi) = W(x)\cos(\phi) + \Delta\phi. \qquad (7)$$

The term $\Delta\phi$ introduces a global tilt (driving torque), while the cosinusoidal term creates local minima (stable traps). The depth of these traps is a function of position: as the oscillators move apart ($|x|$ increases), the potential barriers vanish. The function $V(x, \phi)$ is not conserved but rather acts as a Lyapunov function: the system "slides" down the potential toward a minimum, dissipating energy until it reaches a fixed point. If $\Delta = 0$, the system is a pure gradient system that settles into the nearest local minimum. If $\Delta$ exceeds a certain threshold, the system can no longer reach an equilibrium. It enters a limit cycle (phase drift), where it continuously "descends" the potential slope. This represents an out-of-equilibrium driven system where $\Delta$ acts as a constant energy input. When resting in a local minimum, the system (5)-(6) displays excitability: The system functions as an excitable dipole possessing a stable resting state protected by a threshold—specifically, the separatrix defined by the unstable manifold of the saddle point (see Appendix A). A sub-threshold perturbation allows the trajectory to relax back to equilibrium, but a stimulus exceeding the boundary of the manifold triggers a large-scale excursion through phase space before the system re-stabilizes. Figure 1 illustrates that the system (5)-(6) displays a characteristic response of a Type-I excitable system to external stimuli. Stimuli are obtained by instantly displacing the relative position by a certain amount. Subcritical stimuli (blue traces) induce transient excursions in both the relative spatial and phase, followed by a swift return to one of the stable resting states, in this case: $\phi_{sync} = \pi - \arcsin(\Delta)$ (see Appendix A). In contrast, a supercritical stimulus (red traces) triggers a large amplitude excursion in $x$ and a complete $2\pi$ phase slip, characteristic of an action potential-like response or spiking behavior. The excursion, which includes a phase slip, occurs on a cylindrical manifold. This dynamic results in a screw-like trajectory that involves both the relative position and the relative phase. This phenomenon describes helical excitability, a form of unconventional spatiotemporal excitability that emerges from the coupling of chiral oscillators.

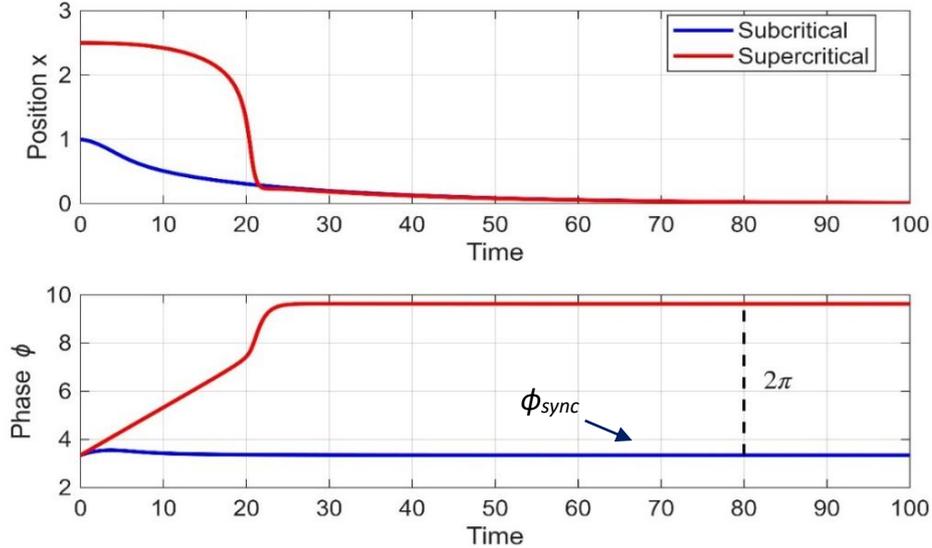

**Figure 1**. Excitability of a Helical Excitable Dipole for $\Delta = 0.2$. The top panel illustrates the time evolution of the relative position, while the bottom panel shows the relative phase. Trajectories for both subcritical (blue lines) and supercritical (red lines) stimuli are displayed. A subcritical perturbation allows the system to relax back to its stable equilibrium, whereas a supercritical stimulus triggers a large helical excursion that corresponds to a full $2\pi$ phase slip.

The helical nature of the excitability is demonstrated by its non-zero helicity density. In the $(x,\phi)$ phase space, helicity density is represented by the closed loop path integral

$$H = \oint x\, d\phi, \qquad (8)$$

this expression defines the Hannay angle [26] of the system, capturing the geometric phase shift that results from the non-conservative winding of the phase. The Hannay angle is a classical analogue of the geometric Berry phase. In mechanical terms, the integral (8) quantifies the net work performed on the spatial coordinate during a full phase rotation. By Green's theorem, this value corresponds to the area enclosed by the trajectory in the $(x,\phi)$ plane. While symmetric responses result in zero helicity (H = 0), a non-zero integral over one excitation cycle identifies the unit as a helical pump, where internal phase rotations are converted into directed spatial motion. Concerning Fig. 1, the numerical evaluation of Eq. (8) yielded H = 2.36 for the sub-critical stimulus and H = 10.79 for the super-critical stimulus, illustrating a kind of excitability that is not merely a temporal 'firing' event; it is instead a coherent maneuver that spatially relocates the dipole together with the phase slip. This arises from the coupling of two units with opposite intrinsic chirality according to Eqs. (5)-(6), so that their relative phase $\phi$ becomes inextricably linked to their spatial separation $x$. While chirality is a property of the individual unit's rotation, helicity is an emergent property of their interaction.

The HED operates as a driven-dissipative actuator where $\Delta$ provides a constant energy bias. A suprathreshold stimulus triggers a non-linear response, forcing the dipole through a helical excursion. Although the spatial coordinate $x$ returns to its initial value, the cycle is non-reciprocal, as confirmed by the finite area enclosed in the $(x,\phi)$ plane (Eq. 8) by the helical pulse. This process constitutes a geometry-dependent stroke where the phase-slip couples to a transient spatial deformation, thus HED functions as a local engine that rectifies the stimulus into a structured helical maneuver.

A further feature of the HED dynamics is the spontaneous symmetry breaking (SSB) of the translational direction along the $x$-axis. Although the underlying potential $V(x,\phi)$ is symmetric with respect to $x$, the excitation process forces the system to select a specific direction for its spatial relocation. This choice is dictated by the phase-space position at the moment of the threshold crossing: a random fluctuation in $\phi$ would break the $x$-parity, coupling the $2\pi$ phase-slip to a directed displacement. Consequently, the HED transforms a scalar energy input $\Delta$ into a vector transport event, establishing the basis for directed active motion in the collective regime.

## 3. Two-dimensional collective dynamics

Exporting the HED to the many-body context introduces a further modeling degree of freedom in how interaction symmetries are defined. This choice is a critical factor because different rulesets generate consistently different outcomes. Specifically, the mutual influence between same-handedness COs must be defined. In this work, I adopt a complementary interaction scheme consistent with previous studies [27-29]. Unlike standard Kuramoto models where all units tend toward global alignment, this scheme introduces a competition between in-phase and anti-phase synchronization dictated by relative chirality. This design fosters a frustrated landscape and aims to enable the emergence of out-of-equilibrium collective modes. Furthermore, a 2D geometry favors the organization into spatially-extended topological structures.

## 3.1 The model

The picture of a many-body CO system is sketched in Fig. 2a; CW and CCW COs are represented by blue and red arrows, respectively, able to rotate and move in $\mathbb{R}^2$, with open boundary conditions. This representation proved more effective than the previous one [27,28,29] for visualizing the morphology of the emerging orientational patterns (defects) and waves, as further elucidated in the manuscript. As mentioned above, the interaction between these chiral units follows a complementary scheme (Fig. 2b) that establishes a competition between in-phase and out-of-phase coordination: Like-Chirality Interaction: Units of the same color repel each other spatially but attract orientationally (favoring in-phase synchronization); Opposite-Chirality Interaction: Units of different colors attract spatially but repel orientationally (favoring out-of-phase synchronization).

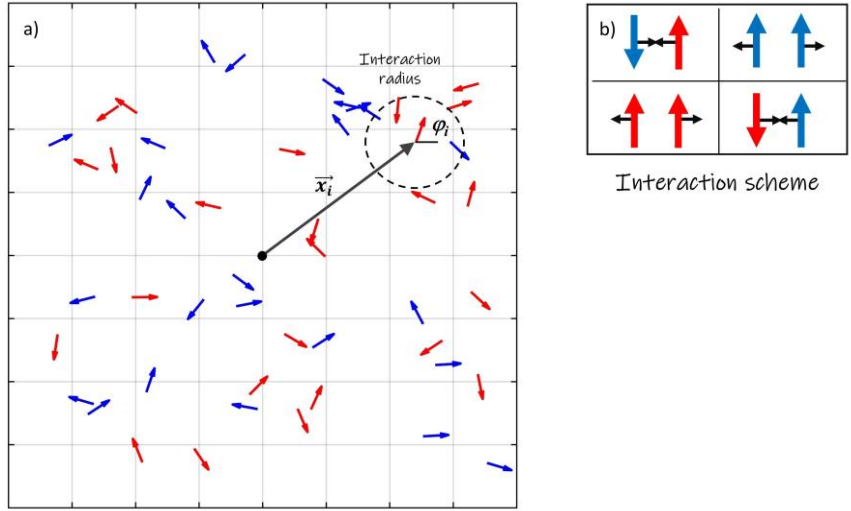

**Figure 2**. Sketch of the Complementary-COs Model. a) The model assigns both a position and an angle in [0,2π] to each CO. b) Interaction forces are complementary based on color: same-colored COs repel positionally and attract orientationally, while different-colored COs do the opposite. All interaction strengths decrease with increasing positional distance.

Each CO is given a positional ($\vec{x}_i$) and an angular/orientational ($\varphi_i$) degree of freedom, with $i = 1…N$, being $N$ the total amount of COs. The equations of motion for each $i$-th CO are

$$\begin{cases} \dot{\vec{x}}_i = \sum_{j=1}^{N} \nabla_i \, W(|\vec{x}_i - \vec{x}_j|) \cos(\varphi_i - \varphi_j), & (9) \\ \dot{\varphi}_i = \gamma_i \Delta + \sum_{j=1}^{N} \gamma_i \gamma_j \, W(|\vec{x}_i - \vec{x}_j|) \sin(\varphi_i - \varphi_j), & (10) \end{cases}$$

where $\nabla_i$ means differentiation respect to the $i$-th direction, $\vec{x}_i$ and $\varphi_i$ are respectively the spatial (in $\mathbb{R}^2$) and angular (in $S^1$) coordinates of the $i$-th unit, $i = 1,…N$, and $|\vec{x}_i - \vec{x}_j|$ is the Euclidean distance in $\mathbb{R}^2$. The functional wells (characteristic length L = 1) are the many-body corresponding of the many-body version of the Gaussian kernel $W(x)$ described in the previous section:

$$W(|\vec{x}_i - \vec{x}_j|) = -e^{-|\vec{x}_i - \vec{x}_j|^2}. \tag{11}$$

The two-valued coefficients $\gamma_i$ define the *i-th* oscillator chirality, i.e., $\gamma_i$ = +1 if the oscillator is a CCW (red) or $\gamma_i$ = −1 if is CW (blue). If N = 2 and one CO is red and one is blue, Eqs. (9)-(10) revert to Eqs. (5)-(6). Mathematically, the model (9)-(10) is a many-body dissipative and non-linear dynamical system, and Δ is the control parameter. Consistently with the previous section, the effect of Δ in Eq. (10) is to split the natural frequencies of red and blue COs, providing a rotational speed term that is clockwise for red COs and counterclockwise for blue COs; in a physical system Δ would be called a circular anisotropy, hence Δ will be referred to as anisotropy in the rest of the manuscript. Collectively, the parameter Δ is expected to trigger an order-to-disorder phase transition in the system upon overcoming a specific collective locking/unlocking threshold (similarly to the Kuramoto model), driving the system to deterministic chaos. As introduced in the previous section, the parameter Δ makes the system active, because it provides the COs a phase drift which, due to variables coupling, can activate positional dynamics. The term $\cos(\phi_i - \phi_j)$ acts as "spin-orbit" coupling—it links the spatial motion to the internal orientation state; this is crucial for topological structure formation as discussed further in the manuscript. The $\sin(\phi_i - \phi_j)$ term acts as the synchronization torque, a hallmark of the Kuramoto model.

Concerns might arise regarding the perceived arbitrariness in the choice of the interaction kernel *W*. However, while the quantitative details of the interaction depend on its exact form, the key qualitative results reported in this work are robust against alternative choices of short-range interaction functions. The use of a Gaussian kernel represents the simplest and most standard choice for capturing the results contained in this manuscript, while also ensuring numerical stability.

*3.2 The isotropic limit and the emergence of order for weak anisotropy.*
In the isotropic case, where Δ = 0, the system behaves as a purely dissipative gradient system. This regime is highly degenerate: the dynamics are dominated by the initial conditions, leading to the formation of branched patterns of phase locked HEDs. In this state, the system becomes trapped in a multitude of local minima within a rugged potential landscape, exhibiting what can be described as residual entropy. These configurations lack long-range order and are reminiscent of glassy or frozen states, where the absence of internal drive prevents the system from escaping metastable clusters. While these instances of residual entropy are scientifically interesting, their detailed analysis is beyond the scope of this work.
However, the introduction of a small anisotropy Δ alters this landscape: the system breaks away from its dependence on initial positional configurations and organizes into crystalline solutions. In this regime, the competition between in-phase and out-of-phase coordination, enforced by the complementary interaction scheme, stabilizes periodic lattices, finite size effects notwithstanding. Importantly, the system still undergoes a Spontaneous Symmetry Breaking (SSB) of the circular phase space, forcing all HEDs to align at a specific, uniform angle that is randomly determined by the initial phase conditions. This phenomenon is shown by the following numerical results: I integrated the equations of motion for N = 80 chiral oscillators with spatial degrees of freedom, using a standard integration method. The anisotropy parameter was set to Δ = 0.1, a value well within the locking range for a single HED. The initial conditions were random: oscillator phases were uniformly distributed in [0, 2π], and positions were randomly distributed within a square box with an initial density of ϱ =1. This density choice is important but not critical, provided the units are not too sparse (preventing clustering) or too compressed (hindering collective flow). The resulting spatiotemporal dynamics (shown in Movie S1) reveal a transition from initial disorder to a highly ordered configuration (reported in Fig. 3). The pattern presented in Figure 3 consists of quasi-regularly spaced HEDs, where each HED satisfies the local stability properties of the single-dipole case. This configuration is hereafter referred to as a Type 1 solution. The spatial regularity is maintained by a repulsive dipole-dipole force, which increases with Δ and decreases with the inter-dipolar distance (further detailed in Appendix B), pushing the pattern to expand at an exponentially vanishing velocity. This is a collective state where no transport (no dynamics) is present (an "insulating" state) and where all HEDs behave as one, i.e., this collective state is described by a well-defined global phase (chosen by SSB), which is uniform across the entire pattern.

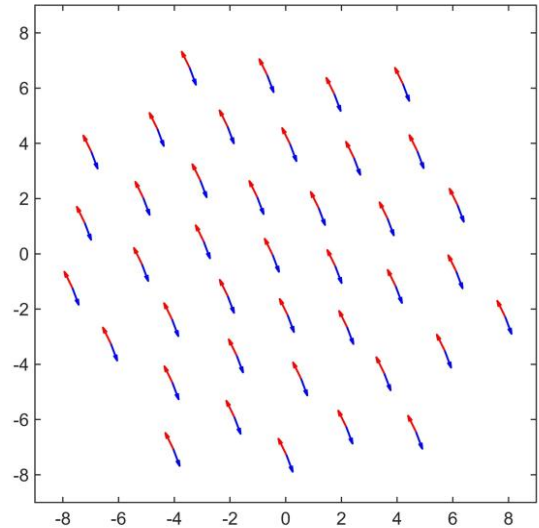

**Figure 3.** A Type 1 solution for N = 80 and Δ = 0.1: A crystalline pattern of Helical Excitable Dipoles (HEDs) possessing both positional and orientational order. This configuration represents a stable equilibrium state resulting from the spontaneous breaking of the continuous circular symmetry. The HEDs are locked into a specific, uniform orientation, determined randomly by the initial conditions.

*3.3 Collective dynamics for moderate anisotropy*

Increasing Δ without crossing the local stability threshold, a different collective behavior shows up, bistable with Type 1. The [Movie S2](#) shows the spatiotemporal evolution of N = 80 COs with Δ = 0.2. After the transient has expired, the system develops a topological vortex complex, which I have termed the Type 2 solution. This configuration exhibits a peculiar twofold structure: a static texture of HEDs curling around a central core, and surrounding this, a dynamic state made of two trains of equidistant COs traveling in opposite directions along the same loop trajectory. Within the loop, the positional and orientational motions are synchronized in a "phase-momentum locked" state, i.e., one full phase rotation ($2\pi$) corresponds to a net spatial translation along the closed path for each CO. In the loop, on average, the distance between adjacent COs of the same chirality is maintained at 1, with minimal deviation. Surprisingly, the red and blue CO trains slide over each other almost "frictionless", despite the value of Δ that would imply attraction and mutual locking in the same position by COs of different chirality.

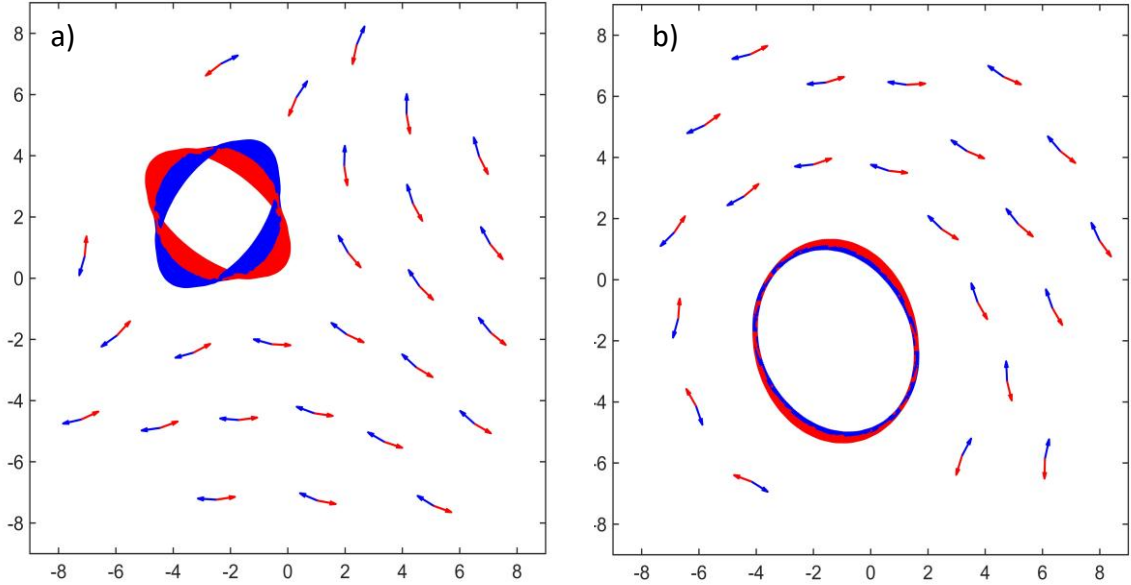

**Figure 4**: Illustrates persistent features by superimposing video frames over time for N = 80 and Δ = 0.2. (a) A stable antivortex complex with a topological charge Q = –1; (b) A stable vortex complex with topological charge Q = 1, emerging for the same parameter values.

Emergent Type 2 solutions are topological point defects driven by a phase singularity, representing a dynamic balance between alignment interactions and "spin-orbit" (phase-momentum) coupling. They appear in two distinct forms (vortex and antivortex) for the same parameter values, as illustrated in a superimposition of video frames (akin to a long-exposure photograph) shown in Figure 4. These forms are classified based on their winding number (or topological charge) Q which measures the net rotation of the COs' orientation along a CCW closed path around the defect core. In both cases (vortex and antivortex), the red and blue CO phase of the static texture rotate together (in the same direction) following a closed path around the defect core; however, within the dynamics loop, their own spatial rotation can be either concordant or discordant with their phase rotation. Specifically, for the stable vortex complex Q = +1, the spatial rotation is concordant with the phase rotation, resulting in a smooth loop trajectory (Figure 4b). Conversely, for the stable antivortex complex Q = -1, the spatial rotation is discordant with the phase rotation in the loop, which produces the characteristic twisted loop shape observed in the Figure 4a. The selection between the vortex complex (Q = +1) and the antivortex complex (Q = -1) is again the result of spontaneous symmetry breaking, i.e. it is determined by subtle differences in the initial conditions.

The attractors emerging from repeated simulations of Eqs. (5)-(6), for moderate anisotropy (Δ between 0.2 and 0.4) and for a given system size (N = 80), show a wide range of morphological diversity; still, all can be ascribed to one out of three distinct states based on their topological properties: Type 1 solutions (equilibrium states with Q = 0), Type 2 vortex (Fig. 5a, d and e) with Q = +1, or Type 2 antivortex (Fig. 5b and f) with Q = -1. The emerging Type 2 solutions display a high degree of variability in terms of: the number of elements involved in the internal loop; the phase singularity position in the plane; the vorticity and the velocity of the transport motion (momentum); and the phase rotation velocity in the loops. Rarely, a double loop with a daughter loop branching off from the main formation (see Fig. 5d) appears, providing evidence of the system's capacity for morphological complexity, as discussed in a later section; however, even that "strange" configuration belongs to the Q = +1 topological class.

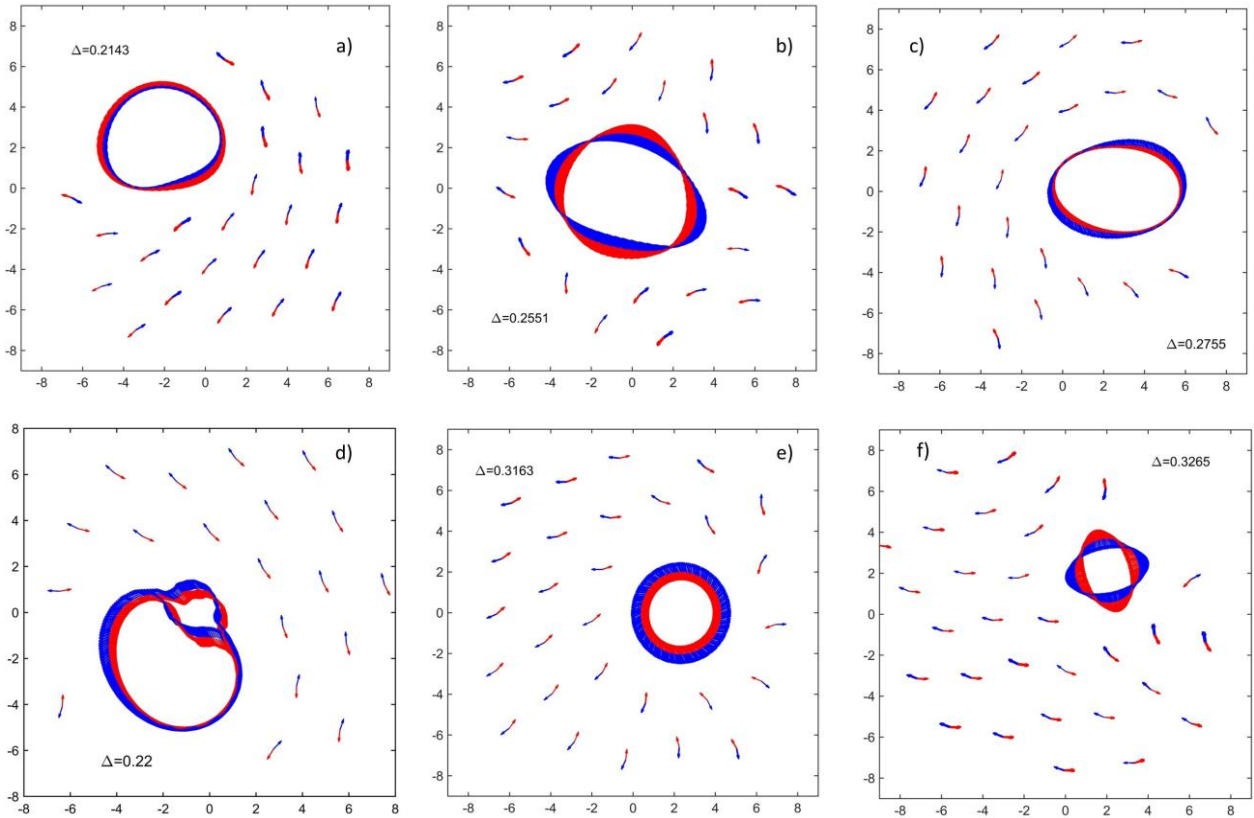

**Figure 5**. Illustrates the diversity of emerging attractors obtained from repeated numerical simulations using different values of Δ, N = 80. The system exhibits multistability, settling into either equilibrium Type 1 solutions or Type 2 solutions, which include both vortices (panels a, d, e) and antivortices (panels b, f) complexes. Type 2 solutions display high variability in the number of elements, position, vorticity, momentum and phase velocity.

Increasing further the anisotropy, numerical simulations for Δ ≈ 0.4 (N = 80) show the emergence of "spurious" clusters of fast rotating and spatially confined COs, in shapes of small rods (see Figure 7) within the Type 2 state (i.e. in compresence with vortices or antivortices). Small rods rotation excites phase vibrations in the neighboring locked HEDs and affects the phase-momentum coherence in the vortices. This highlights a complex interplay between localized rotational motion and the collective, large-scale order of the system.

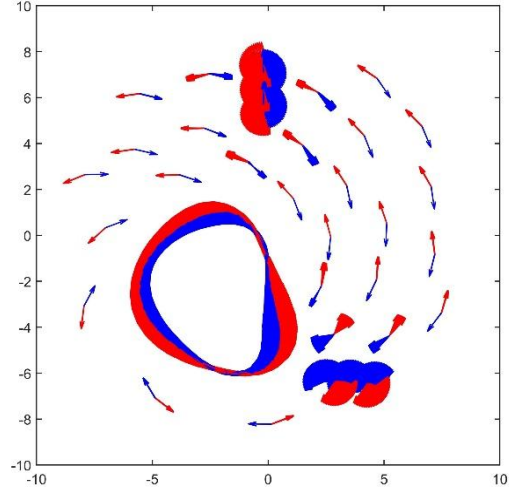

**Figure 6:** Emergence of spatially confined, fast-rotating chiral oscillators (COs) clusters for Δ = 0.4, N = 80 within a Type 2 state. The vector field illustrates the high rotational velocity of the CO rods (larger, curved arrows), which excite phase vibrations in the surrounding phase-locked HEDs.

Increasing Δ further to approximately 0.5, the system exhibits spatiotemporal chaotic dynamics in both position and orientation, a form of active turbulence. I refer to these dynamic states as Type 3. Further increasing Δ, the behavior becomes more erratic, and eventually no trace of order remains.

### *3.4 Robustness against perturbations*

The primary implication of a topological invariant is robustness. In a topological state, the value of the invariant remains unchanged under continuous deformations, strong perturbations, or a high degree of diversity in spatial arrangements. In fact, the complementary regime structures obtained in the previous subsection, made of vortex and antivortex complexes, proved to be highly resilient with respect to different kinds of perturbations. First, the structures persist stably under the addition of Langevin white noise terms to Eqs. (9)-(10) even at high levels of noise (a more detailed study will be the subject of future investigations). Second, I perturbed the structure that emerged in Movie S2 in the following way: after the structure is formed, I removed one red CO from a HED forming the static texture and placed it into the loop, close to the center. The resulting dynamics are reported in Movie S3, which shows that the displaced red element is quickly absorbed into the vortex and, after a short time, a red CO is expelled from the internal loop so that it can couple back with the uncoupled blue CO, and reform a HED to restore the original topology. This observed self-repairing capability is a direct consequence of the non-local topological protection inherent to these complexes. The reported results suggest that a relatively small ensemble of N = 80 COs is able to display a rich multi-stable scenario that requires a systematic investigation.

## 3.5. Statistical analysis

Considering the system (9)-(10), I have performed reiterated numerical simulations with $\Delta$ spanning from 0 to 1, and with different realizations for each value of $\Delta$, starting from already discussed starting conditions. First, I have used the averaged kinetic 'temperature' $T$ as a reference, which is a measure of the system's average spatial activity

$$T = \frac{1}{N}\langle \sum_{i=1}^{N} \dot{x}_i^2 \rangle, \qquad (7)$$

where $\langle \cdot \rangle$ denotes the time average. The second parameter describes the rhythm of the collective phase dynamics. By transposing to many-body the phase difference $\phi$ employed in the previous Section, I calculate the mean folded phase which accounts for the chirality $\gamma$ of each oscillator

$$\Phi = \frac{1}{N}\langle \sum_{i=1}^{N} \gamma_i \varphi_i \rangle, \qquad (8)$$

which, for N = 2 chiral oscillators of opposite chirality, reverts (up to a factor 2) to the phase difference $\phi$ of a single HED, discussed in Section 2 and in Appendix A.
The associated collective period $\tau$ is consequently defined as:

$$\tau = \frac{2\pi}{\Phi}, \qquad (9)$$

which reverts to the oscillation period introduced in Appendix A for a single HED. The results for $T$ versus $\Delta$ are shown in Fig. 7a.

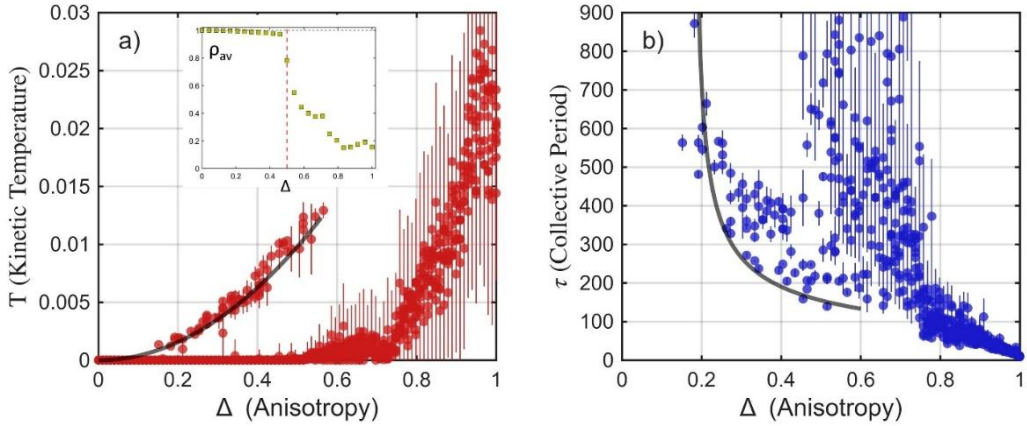

**Figure 7**. The data are obtained from numerical simulations: 10 distinct realizations for each value of $\Delta$. Panel (a) shows the scatter plot of the time-averaged kinetic temperature $T$ as a function of the control parameter $\Delta$; the plot reveals distinct behavioral regimes: a small $\Delta$ regime (Type 1 patterns) with vanishing $T$; a bistable intermediate regime $\Delta_{low} < \Delta < \Delta_{high}$ ($\Delta_{high} \approx 0.58$, $\Delta_{low} \approx 0.18$) featuring a lower branch of Type 1 patterns and an upper branch of Type 2 patterns (fitting with $T = a\Delta^2$, $a = 0.04$); and a supercritical, turbulent, chaotic regime for $\Delta > \Delta_{high}$. Inset in panel (a) displays the time-averaged absolute value of the Kuramoto order parameter ($\rho_{av}$) as a function of $\Delta$. The panel illustrates a Kuramoto-like phase transition behavior, specifically the transition from Type 1 (static phase synchronization, $\rho_{av} = 1$) to Type 3 (desynchronization/turbulence, $\rho_{av} \to 0$), obtained from initial conditions near the Type 1 basin of attraction. Panel (b) pictures the collective period $\tau$ (mean and standard deviation) versus $\Delta$. Fitting (partially) with $\tau = C/\sqrt{\Delta - \Delta_{low}}$ with C = 119)

Panel (a) of Fig. 7, shows $T \approx 0$ for low values of $\Delta$, approximately $0 < \Delta < 0.18$. In this region the equilibrium Type 1 patterns appear, stably. For $0.18 < \Delta < 0.58$ the values of $T$ separate in two branches. The lower branch is a continuation of the Type 1 pattern branch from the previous interval, and is bistable with a second branch containing the twofold Type 2 patterns discussed in the previous subsection. The upper branch follows approximately a $\Delta^2$ dependence (grey fitting line in Fig.6a) and contains an infinite number of different attractors, separated in two families classified via the topological charge Q). For $\Delta > 0.58$, local instabilities take place and all the COs unlock destroying both Type 1 and Type 2 attractors, and chaotic dynamics (Type 3 attractors) appear, increasingly turbulent as $\Delta$ increases further and any trace of order is finally lost. The values of $\Delta$ = 0.58 for local instabilities consistently differs from the instability threshold of a single HED ($\Delta = 1$), this is so because at moderate $\Delta$ values, local instabilities and spatially confined, fast-rotating clusters begin to appear in the collective regime, perturbing large-scale coherence and driving the system towards chaotic dynamics (Type 3) before the single HED reaches its instability threshold of $\Delta = 1$. The local instability threshold of the collective system is therefore reduced due to these emergent many-body dynamics, which are not predicted by the single-dipole model. Current coarse-grained analyses are underway to better quantify this specific aspect.

The transition from Type 1 to Type 3 patterns is a "classical" spontaneous symmetry breaking phase transition of the Landau type, understood in terms of the breaking of a local circular symmetry, parallel to the transition to a ferromagnetic state [30] or the Kuramoto transition to collective synchronization. It can be described using the Kuramoto order parameter, modified as in previous works [26-28]:

$$\rho e^{i\theta} = \frac{1}{N}\sum_{k=1}^{N} \gamma_k e^{i\varphi_k}. \qquad (10)$$

The time averaged absolute value $\rho_{av} = \langle\rho\rangle$ is bounded between 0 and 1, representing total phase desynchronization and total phase synchronization, respectively. The inset in Panel (a) of Fig. 7 shows the typical Kuramoto-like phase transition behavior when $\rho_{av}$ is computed versus $\Delta$ as it concerns the transition Type 1 to Type 3, obtained by making the system start from initial conditions close to the basin of attraction of Type 1 solution. The transition occurs at approximatively $\Delta=0.58$, confirming this value as the threshold for local unlocking and chaos.

The kinetic temperature $T$ data for the upper branch of Type 2 solutions, presented in Figure 7a, (partially) follows a quadratic dependence:

$$T = a\Delta^2, \qquad (11)$$

with $a$ = 0.04 as a fitting parameter. The origin of this $\Delta^2$ behavior is currently under investigation.

Panel (b) of Fig. 7 displays the collective period $\tau$ (mean and standard deviation) as a function of $\Delta$. The data can be(partially) fitted by

$$\tau = C/\sqrt{\Delta - \Delta_{\text{low}}}, \qquad (12)$$

with $C$=119 is a fitting parameter and $\Delta_{\text{low}}$ = 0.18, as far as Type 2 attractors are concerned. The behavior of the collective period $\tau$ in Fig. 7b suggests the presence of a global Saddle-Node bifurcation acting as the organizing center for the emergence of the Type 2 patterns. The (partial) fitting (12) of the collective period $\tau$ is the signature of critical slowing down that generically occurs as a control parameter approaches the Saddle-Node bifurcation point (here $\Delta_{low}$). This provides the mechanism for the system to access a new class of collective states that exhibit topological features. I name those states *superexcitable*, for their collective, coherent, and dissipationless nature, and for emerging from the global version of the same kind of bifurcation (Saddle-Node) that governs conventional type I excitability. Additionally, superexcitability takes place within the locally excitable range of the control parameter. This means the individual COs are not simply being dragged above their local excitability threshold; instead, the collective interaction itself creates a new, global bifurcation within the locally locked region that involves a significant restructuring of the system's entire phase space topology. The states (Type 2 patterns containing stable vortex and antivortex complexes) that emerge from this transition display characteristics typically associated with topological matter: robustness, non-local character, quantized invariants, and phase-momentum locking. In order to quantify the phase-momentum locking I created a global time-averaged parameter $S$, defined as

$$S = \langle \sum_j (\boldsymbol{v}_j e^{i\phi_j}) \rangle, \qquad (13)$$

where the velocity (linear momentum) within definition (13) is introduced through an artificial complex number $\boldsymbol{v} = v_x + iv_y$, a construct designed to facilitate the interaction and coupling between the momentum angle and the intrinsic phase of each unit. The parameter S acts as an effective complex order parameter for identifying the onset of phase-momentum locking within the global dynamics, in practice it serves as a measure of the instantaneous, zero-lag cross-correlation between the linear momentum vector and the intrinsic phase orientation of each chiral oscillator. A high value for |S| indicates a high degree of coherence, analogous to the emergence of collective order in classical statistical mechanics models, such as the magnetization in the Ising model below its critical temperature [30]. More specifically, parameter S serves as the macroscopic, mean-field analogue of a coherent state found in quantum mechanics and optics, representing the emergence of a stable amplitude dynamics and a well-defined collective phase. This behavior draws a parallel to the complex order parameter $\Psi$, that describes phase transitions to the superfluid or superconducting state (in Ginzburg-Landau (GL) theory [31]) where $|\Psi|^2$ quantifies the condensate density and topological defects (like quantized vortices) are characterized by phase winding. However, here S acts as the order parameter for a topological phase which describes a fundamentally different scenario with respect to GL. The observation of non-zero | S | separates coherent phase-momentum locked states (vortices - Type 2 states) from both static Type 1 patterns and from disorganized, high-variance Type 3 states.

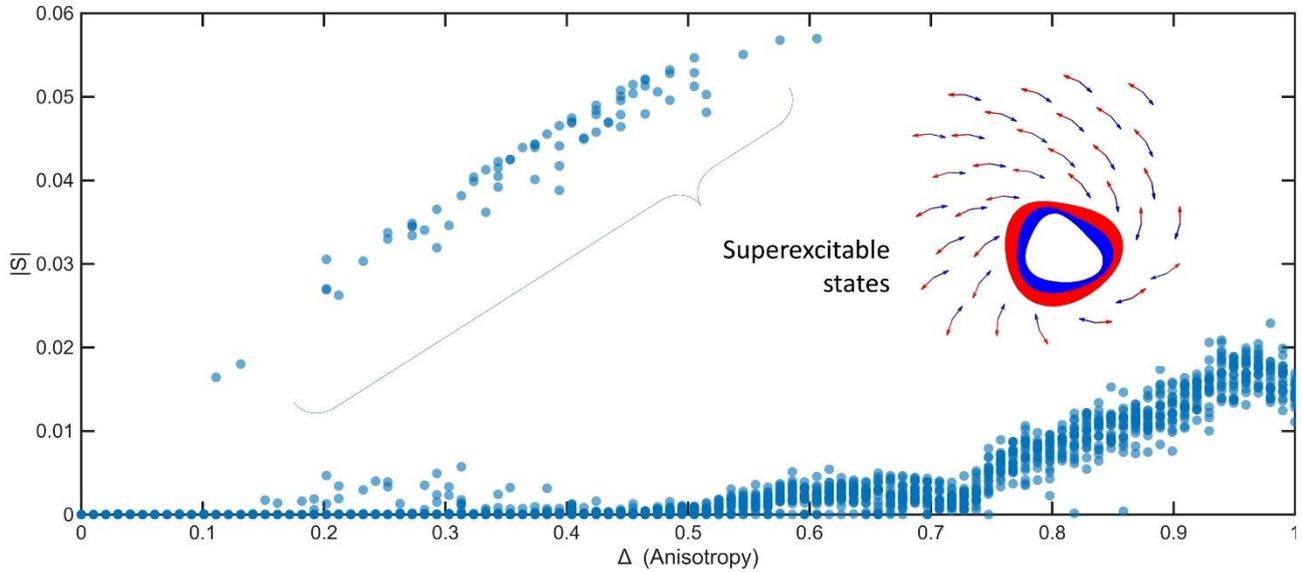

**Figure 8.** Representation of the phase-momentum locking order parameter |S| as a function of the anisotropy Δ. Data points from 20 distinct realizations for each value of Δ are shown. The points with high |S| highlight the superexcitable states (Type 2 solutions), characterized by coherent dynamics. An example of a "clean" superexcitable state is reported in the inset for Δ = 0.3.

The data related to the kinetic temperature $T$ and the collective period $\tau$ (Fig. 8) provide an overview of the phase diagram, highlighting the regions of activity and oscillation frequency. However, these parameters show two partially superimposed transitions: the superexcitable and the Kuramoto-like. Consequently, the localization of the superexcitable phase is less clear, as it is partly hidden behind "dirty" or metastable states characterized by non-optimal phase-momentum coherence. The order parameter S serves as a specific theoretical tool to characterize the superexcitable transition, isolating the "clean" vortices (see inset of Fig.8) and antivortices, because the observation of a non-zero |S| value separates the coherent states with phase-momentum locking (Type 2) from both the static Type 1 patterns and the disorganized Type 3 states.

In essence, the Kuramoto parameter $\varrho_{av}$ and the novel parameter S distinctly characterize the two different transitions discussed above. Specifically, $\varrho_{av}$ captures the breaking of the local circular symmetry associated with a standard Kuramoto-like transition, while S quantifies the non-local, topological order linked to the superexcitable states.

### 3.6. Exploring Larger systems

The results obtained in the previous section concern a sample system size of N = 80. While this might appear a relatively small number for large-scale statistical analysis, I found that for larger sizes the scenario becomes significantly more complex, exhibiting irregularly long "glassy" transients, that are resistant to analysis in this statistical framework. This choice (N = 80) was deliberate to allow for a preliminary characterization of the collective dynamical mechanisms and the resulting phase diagram, acknowledging potential finite-size effects on the precise transition points.

In this subsection, the evolution of superexcitable states is examined as the system size N increases by means of some examples. As a general remark, the single loop patterns observed in smaller systems are replaced by a flexible, morphogenetic flow that generates dynamic compartments and progressively more intricate phase-momentum locked flows as N increases, displaying both long-range order and topological protection. Figure 9 illustrates the progression from: panel (a) for N=200 (simple figure-eight shape, see Movie S4 for the regime spatiotemporal dynamics) to panel (b) for N=500 (more convoluted boundary loops, see Movie S5) to panel (c) for N=1000 (highly intricate, compartmentalized network, see Movie S6). A detailed analysis of the N = 1000, $\Delta$ = 0.01 case is provided in a previous work [29]. The phase-momentum relationship is still a kind of synchronization, but more intricate with respect to the simple 1:1 locking displayed by smaller systems.

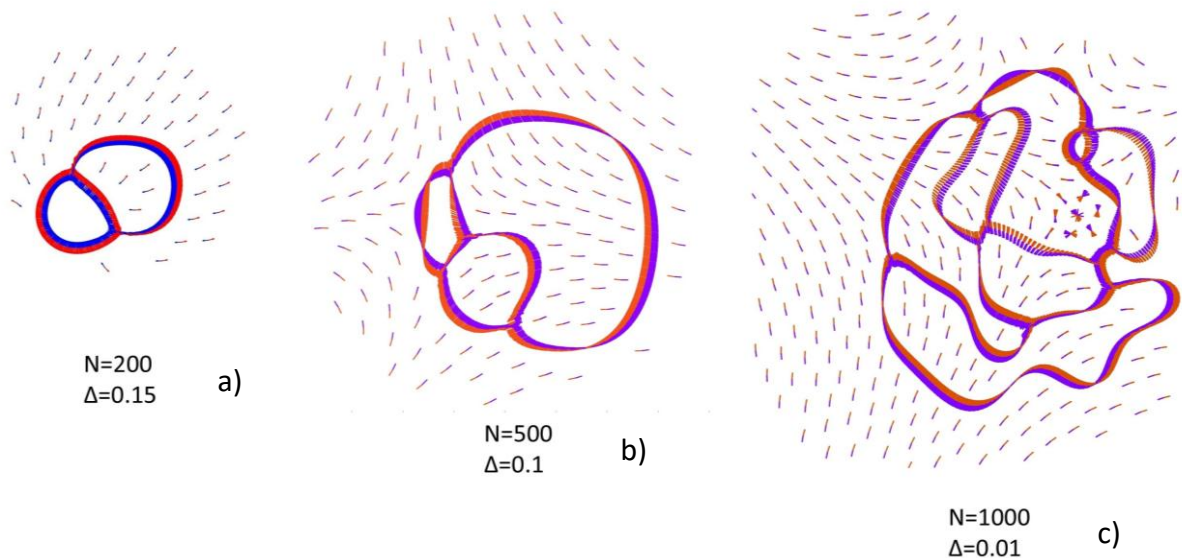

**Figure 9**. Sketch illustrating some examples of the increase in complexity of superexcitable states, as system size N increases. The single vortex observed in small systems is replaced by a progressively complex, morphogenetic flow that forms increasingly intricate phase-momentum locked networks.

Fig. 9c shows instabilities, which indicate a weakening of the topological protection. Indeed, in [29] it is reported that such network is only marginally stable and has a finite lifetime; a possible reason for it is discussed in the next section.

The "big sizes" are numerically hard to tackle; the flow stabilizes after a long, non-exponential "glassy" transients. Many scenarios open, the mathematical explanation of the process that governs such complex patters will be the object of future research.

*3.6. Robustness to phase randomization and topological jumps*

In this subsection, I investigate the stability of a complex superexcitable state for a system size of N = 500. As illustrated in Fig. 8(b), the such state is characterized by a complex topological defect—specifically an anti-vortex—carrying a topological charge of Q = -1. To test the resilience of this state, I introduce a massive stochastic perturbation by randomizing all phases at a given time (see [Movie S7](#)). Following this perturbation, the system does not relax back to its original state; instead, it undergoes a rapid structural reorganization, jumping to a new steady-state configuration. This new regime is characterized by a higher-order topological defect with a charge of Q = -2 (see Fig. 10).

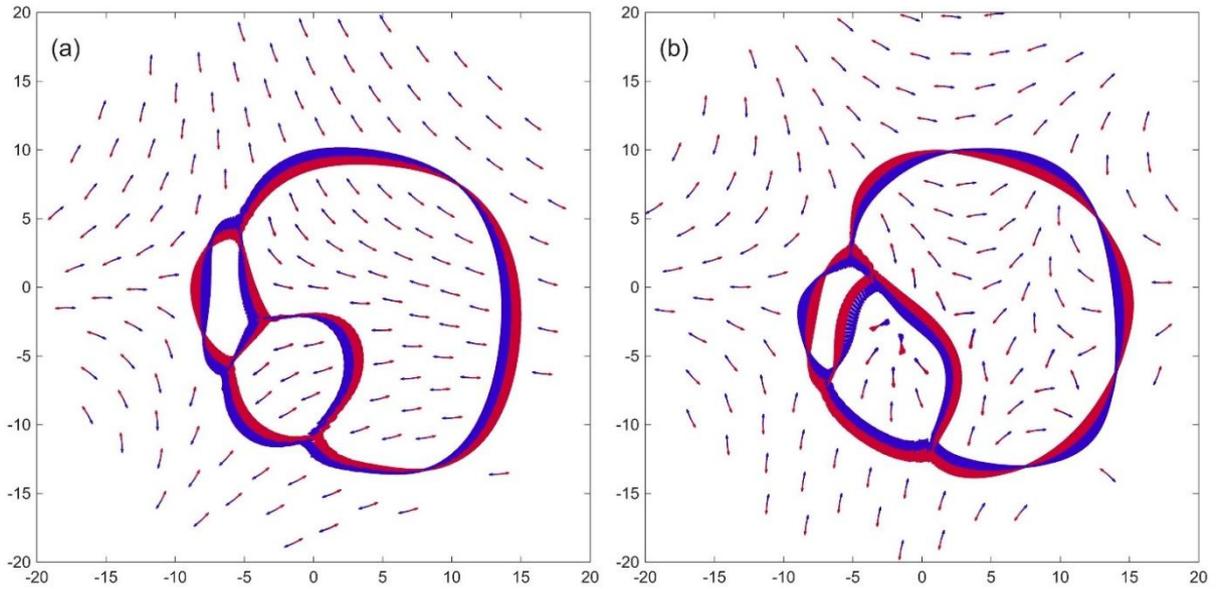

**Figure 10.** Quiver plots of the phase-momentum locked transport regime for N=500. (a) Vector field configuration in the 2D plane before the global phase perturbation, this panel is a reproduction of Figure 9(b), repeated here for direct comparison purposes. The system is in a complex, stable, superexcitable state with a topological charge of Q = - 1. The image is an overlap of multiple final frames, illustrating the static nature of the particle distribution. (b) The system's new stable configuration after an abrupt randomization of all phases. The system relaxes into a new topological sector with a quantized charge of Q = - 2.

The strong perturbation—which completely destroys the coherence of the structure and alters the transport coordination—does not annihilate the existing order. Instead, it forces a topological jump corresponding to a discrete change to a higher integer value of the topological charge. This behavior, characterized by the system seeking an alternative stable topological sector, is a distinct signature of a collective topological state. While far from exhaustive, this simple numerical evidence suggests that these complex topological dynamics warrant a more detailed analysis in future work.

*3.7. Phase diagram in the N-Δ plane.*

By exploring different system sizes, ranging from N = 80 to N = 1000, I have observed that the minimum values of Δ required to obtain Type 2 solutions ($\Delta_{low}$) decrease as the system size N increases, whereas the transition to chaos stabilizes around $\Delta_{high} \approx 0.5 - 0.6$. To highlight the transition between the stability region of Type 1 and the bistable region containing both Type 1 and 2, I performed extensive numerical simulations across the (N, Δ) parameter plane. The result is reported in Fig. 11, where white dots indicate the presence of a Type 1 uniform "insulating" solution and black dots denote the emergence of Type 2 superexcitable vortex complexes. The analysis reveals that as N increases, the transition value $\Delta_{low}$ decreases following the empirical scaling relationship

$$\Delta_{low}(N) = \frac{a}{N}, \qquad (11)$$

where $a$ is a fitting parameter. Repeated simulations consistently yielded similar results, and data from single runs at higher values of remained consistent with the scaling behavior described by Eq. (11). As an addendum to previous results, what was called "network death" in [29] can be understood in the light of Eq. (11). In [29], the emergence of a complex phase-momentum locked network with N = 1000 and Δ = 0.01 (sketched Fig.7c) was reported. The network revealed not to be stable in the very long run. In fact, Eq. (11) gives $\Delta_{low}$(N=1000) = 0.012 as the lower stability threshold for those kinds of solutions; so, being close to threshold but outside the stability region, the network was marginally stable and did not endure.

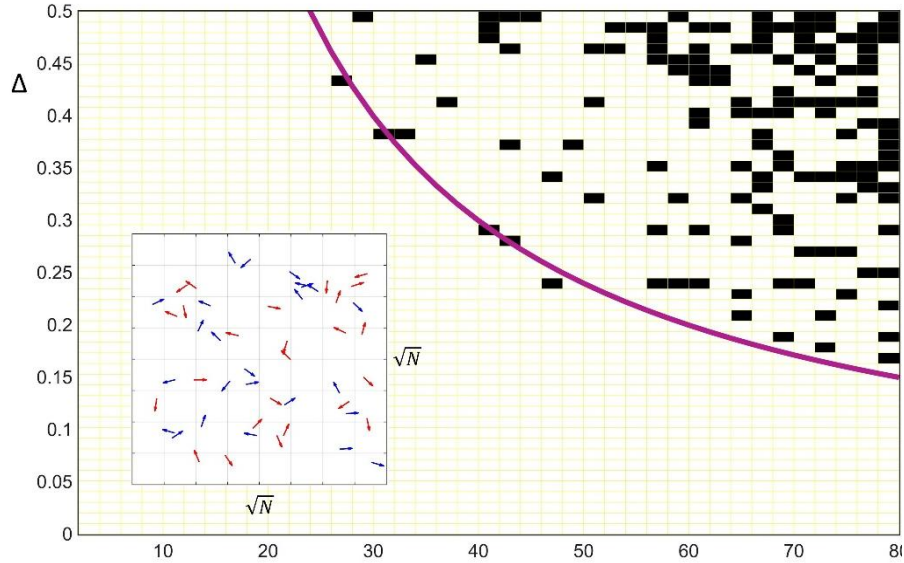

**Figure 11**: Phase diagram in the (N, Δ) parameter space, illustrating the transition between different solution types. White dots indicate regions where stable Type 1 uniform solutions appeared, while black dots denote the emergence of Type 2 complexes at least once in 5 realizations for the same values on N and Δ. The boundary value $\Delta_{low}$ is fitted with $\Delta_{low}(N) = \frac{a}{N}$ (purple curve); $a$ = 12..

Since the density is fixed, N (the number of COs) also represents the area of the squared box where the dynamics initiate (see inset in Fig. 11). Consequently, the significant value is the product of the anisotropy Δ and the system size N. This product N$\Delta_{low}$ effectively acts as a minimum 'quantum' necessary to assist the formation of the superexcitable phase, and, since Δ cannot exceed $\Delta_{high}$ ≈ 0.5 which would activate chaotic behavior, the minimum size for having vortex/antivortex complexes is N ≈ 24. The number N = 24 appears to be the smallest number of COs that can structurally support the complex arrangement required for a stable, self-organized vortex or antivortex composite; it represents the point where the behavior transitions from single-unit interactions to the onset of collective behavior: Below N = 24, the local interactions dominate, and vortices cannot "nucleate" effectively.

## 4. Discussion

This work introduces a framework for studying the reciprocal effects of synchronization dynamics and physical space organization in a system of chiral oscillators (COs), possessing an intrinsic rotational handedness. The model features two populations of these units with antagonistic positional and orientational interactions, which introduce frustration into the system dynamics. The COs are organized in two families (red and blue) with opposite chirality tuned by a control parameter, a form of circular anisotropy. The interaction forces tend to create red/blue locked dipoles, which emerge as excitable structures, hereafter termed Helical Excitable Dipoles (HEDs). HEDs are characterized by a non-zero helicity density represented by the Hannay angle of the system, a classical analogue of the geometric phase. A non-zero integral over one excitation cycle identifies the HED as a helical pump, where internal phase rotations convert into oriented spatial dynamics. This "metric consequence" of excitation distinguishes the HED from classical excitable oscillators and forms the basis for the collective vortices observed in the many-body case.

Though the work spans a range from N=2 to N=1000, it focuses primarily on an intermediate system size N = 80, a challenging domain that straddles the boundary between standard dynamical systems analysis and statistical mechanics. This specific range of dimensions is where the inherent analytical difficulty lies, as the system is too large for exhaustive bifurcation analysis and too small for conventional thermodynamic limits to apply effectively. It is within this intermediate scale that the rich diversity of collective behaviors, often "washed out" in the thermodynamic limit, emerges due to the synergetic cooperation of multiple degrees of freedom. The analytical challenges presented were addressed in this work through the combination of numerical simulations and statistical tools.

The system displays rich topological characteristics in the collective regime. The coupling of orientational and positional degrees of freedom facilitates the existence of point defects characterized by integer winding numbers and associated phase singularities. These are composite patterns of topological orientational textures and phase-momentum locked loops. The flow in these loops is ballistic, which contrasts with predictions based on local properties. These characteristics are hallmarks of a topological phase, and the patterns classify as topological dissipative structures. This phase, which I defined as superexcitable, promotes a non-dissipative collective transport by analogy with certain topological phases of matter, such as those found in topological insulators and superconductors, and constitutes a whole new class of out-of-equilibrium states. A global saddle-node bifurcation acts as the critical switch that allows the system's dynamics to transition into superexcitability, which manifests as robust vortex complexes characterized by phase-momentum locking, paralleling the spin helical transport found in topological insulators [32]. The results indicate that these behaviors emerge well within the local excitability region, suggesting that they are the collective spatio-temporal manifestation of excitability, and not simply the result of drag generated by some suprathreshold oscillators. Essentially, the intrinsic coupling between phase and spatial momentum, combined with the topological protection of the collective structure, creates an extremely efficient and non-dissipative transport mechanism that differs from classical diffusive spatiotemporal excitability [9-10].

The investigation included the numerical evaluation of collective parameters such as kinetic temperature and collective period, along with a Kuramoto order parameter. The system displays a dual phase transition: an "ordinary" symmetry-breaking transition (Kuramoto-type synchronization) coexists with a new topological (superexcitable) transition that activates the vortex complexes, unveiled by a newly introduced collective parameter that emphasizes dynamic states with high degree of phase-momentum locking. The evaluated boundaries of the topological transition in the parameter plane suggest that the topological phase activates when a specific amount of the product between the anisotropy and the system size becomes available.

Increasing the system size leads to greater complexity in collective organization. Unlike the single vortex observed in smaller systems, larger systems show a flexible morphogenetic flow with complex phase-momentum locking patterns.

This work provides a preliminary numerical outlook on the phenomenology of this complex system and proposes collective parameters for identifying the various regimes, thereby paving the way for future, more comprehensive research into these dynamics.

Beyond theoretical considerations, the described scenario finds various applications in biophysics. The Helical Excitable Dipole (HED) model is conceptually similar to a biological ratchet mechanism. Both rely on breaking specific symmetries to achieve a polarized, non-equilibrium response to movement, as occurs in molecular motors operating at scales where thermal fluctuations are significant. The concept of motile excitable units that self-organize into collective, adaptive, and topologically protected rhythms is applicable to the coordinated movement of flagella [22], migrating epithelial tissues [7], and the metachronal coordination of cilia [8], where each beat is a mechanical act involving displacement. In cardiology, during the embryonic phase, the myocardium undergoes a sophisticated self-organization characterized by a transition from a simple cell layer to a complex, pulsating structure. Myocardial progenitor cells exhibit spatial motility, migrating from the mesoderm to coalesce into fields that form the heart [33]. The formation of the helical structure (looping) of the embryonic heart tube, essential for its final function as a pump, is a key example of how intrinsic cellular chirality translates into large-scale deformation and helical movement, a process that ensures the correct orientation of the organ. Excitation in these systems is not merely a temporal event but a topological act of symmetry breaking that translates into net spatial transport. The presented model applies to such scenarios because it simulates how motile excitable units self-organize into a topologically protected collective rhythm, providing a unified framework for understanding collective movement across diverse biological contexts.

**Supplementary Materials:** The following supporting information can be downloaded at:

Video S1: title: Uniform type 1 pattern.
Video S2: title: Antivortex complex.
Video S3: title: Robustness against rupture
Video S4: title: Regime pattern for N=200, Δ=0.15.
Video S5: title: Regime pattern for N=500, Δ=0.1.
Video S6: title: Regime pattern for N=1000, Δ=0.01.
Video S7: title: Response to phase randomization for N=500, Δ=0.1.

**Author Contributions:** A.S. is the sole author of this manuscript and was responsible for all aspects of the reported research, including conception, theoretical development, analysis, and manuscript preparation.

**Funding:** This research received no external funding

**Institutional Review Board Statement:** Not applicable

**Informed Consent Statement:** Not applicable

**Data Availability Statement:** The author confirms that all data supporting the findings of this study are available within the article and its Supplementary Materials

**Acknowledgments:** The author thanks the University of Pavia for institutional support and is grateful to Prof. Pierre-Simon Jouk (Department of Medical Genetics, CHU de Grenoble, and TIMC-IMAG, Université Grenoble Alpes, Grenoble, France) for insightful conversations on cardiac dynamics. Additionally, the author wishes to thank Prof. Luis L. Bonilla (Universidad Carlos III de Madrid, Madrid, Spain) and N. O. Rojas (Université Côte d'Azur, Nice, France) for precious insights on an earlier version of the manuscript.

**Conflicts of Interest:** The author declares no conflicts of interest

**Abbreviations**

The following abbreviations are used in this manuscript:

CO (Chiral Oscillator), HED (Helical Excitable Dipole), CW (ClockWise), CCW (CounterClockWise), SN (Saddle-Node), SSB (Spontaneous Symmetry Breaking).

# Appendix A. Stability analysis of the Helical Excitable Dipole (HED)

The system describes the coupled evolution of a relative phase $\phi$ and a relative spatial coordinate $x$ on a cylindrical manifold $\mathbb{R} \times S^1$. For the reader's convenience, I reproduce here the Equations (5)-(6), relabeled as (A1)-(A2):

$$\begin{cases} \dot{x} = W'(x)\cos(\phi) = 2xe^{-x^2}\cos(\phi) & \text{(A1)} \\ \dot{\phi} = \Delta - W(x)\sin(\phi) = \Delta + e^{-x^2}\sin(\phi) & \text{(A2)} \end{cases}$$

Where $W(x) = \exp(-x^2)$ is the Gaussian interaction kernel (short-range) and $\Delta$ is the intrinsic frequency detuning.

## A.1. Stability analysis of the HED for Δ =0

*Nullclines*

The nullclines are the sets of points in the $(x,\phi)$ phase space where either $\dot{x} = 0$ or $\dot{\phi} = 0$. The $x$-nullcline for Eq. (A1) is

$$2xe^{-x^2}\cos(\phi) = 0 \quad \text{(A3)}$$

This condition is satisfied if: $x = 0$ or $\cos(\phi) = 0$ which means $\phi = \pi/2 + n\pi$ for any integer $n$. The $\phi$-nullcline for Eq. (A2) is

$$e^{-x^2}\sin(\phi) = 0, \quad \text{(A4)}$$

this condition is satisfied if: $\sin(\phi) = 0$, which means $\phi = n\pi$ for any integer $n$.

*Equilibrium Points*

Equilibrium points (fixed points) occur where both nullclines intersect. The equilibrium points are therefore

$$(x_0, \phi_0) = (0, n\pi) \quad \text{(A5)}$$

for any integer $n$.
The Linear Stability Analysis (LSA) indicates that the equilibrium points $(0, 2\pi n)$ are unstable nodes (sources), whereas the equilibrium points $(0, \pi + 2\pi n)$ are stable nodes (sinks).

Even in the isotropic limit where Δ=0, the basin of attraction of the single stable fixed point displays a non-trivial geometry, shown by a numerical investigation. The outcome graph (Fig. A1) illustrates the following elements: the colored background (red and blue) represents the basin of attraction. Each colored area groups initial conditions that converge to a specific long-term behavior. The blue area indicates initial conditions that converge to the stable, locked fixed point (green circle) at $(0, \pi)$. The red area indicates initial conditions whose trajectories do not converge to the locked state, leading instead to unbounded phase rolling.

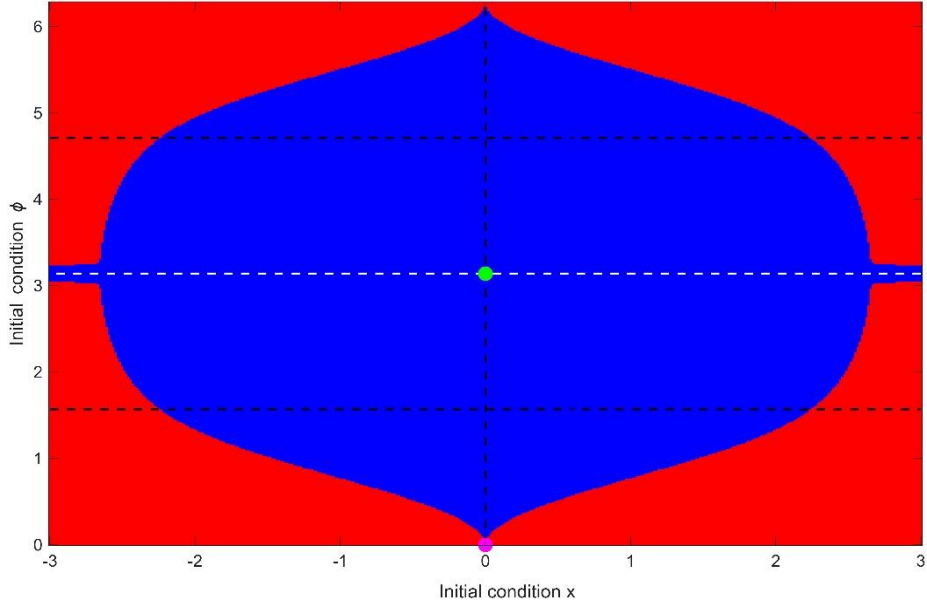

**Figure A1** Basin of attraction map for the Helical Excitable Dipole system in the isotropic, purely dissipative limit Δ=0. The color overlay indicates the final long-term behavior based on initial conditions. The blue region denotes initial conditions that converge to the stable, stationary locked state (green circle). The red region indicates initial conditions that result in unbounded phase rolling. The dashed lines represent the nullclines, white for $\phi$, black for $x$. The fixed points are marked by circles: green for the stable node and magenta for the unstable node.

## A.1.2. Stability analysis of a HED for Δ > 0

### Nullclines

The condition for the x-nullcline remains unchanged from the Δ = 0 case, The condition for the $\phi$ nullcline changes due to Δ

$$\sin(\phi) = -\Delta e^{x^2} \quad (A6)$$

The conditions for the existence of real solutions of Eq. (A6) are $\Delta \leq 1$ (Δ has been taken non-negative).

### Equilibrium Points

Fixed points are the intersections between the nullclines. There are two main families of solutions in the interval $[0, 2\pi)$:

$$\phi_{sync} = \arcsin(-\Delta) + \pi \quad (A7)$$

is the stable/locked state (or *sink*, the green point in Fig. A2), whereas

$$\phi_{unsync} = \arcsin(-\Delta) \quad (A8)$$

is unstable/unlocked (a *source*, the magenta point in Fig. A2).

Differently from the previous $\Delta = 0$ case, here a saddle point is also present, create by the nullcline $\phi = \pi/2 + n\pi$. For $\phi = \pi/2$:

$$1 = -\Delta e^{x^2} \Rightarrow e^{x^2} = -\frac{1}{\Delta} \quad (A9)$$

This solution exists only if $\Delta < 0$ so can be dismissed. For $\phi = 3\pi/2$

$$-1 = -\Delta e^{x^2} \Rightarrow e^{x^2} = \frac{1}{\Delta} \quad (A10)$$

This solution exists only if $\Delta > 0$ and the position $x$ of these saddle points (cyan points in Fig.A2) are

$$x_{saddle} = \pm\sqrt{-\ln(\Delta)} \quad (A11)$$

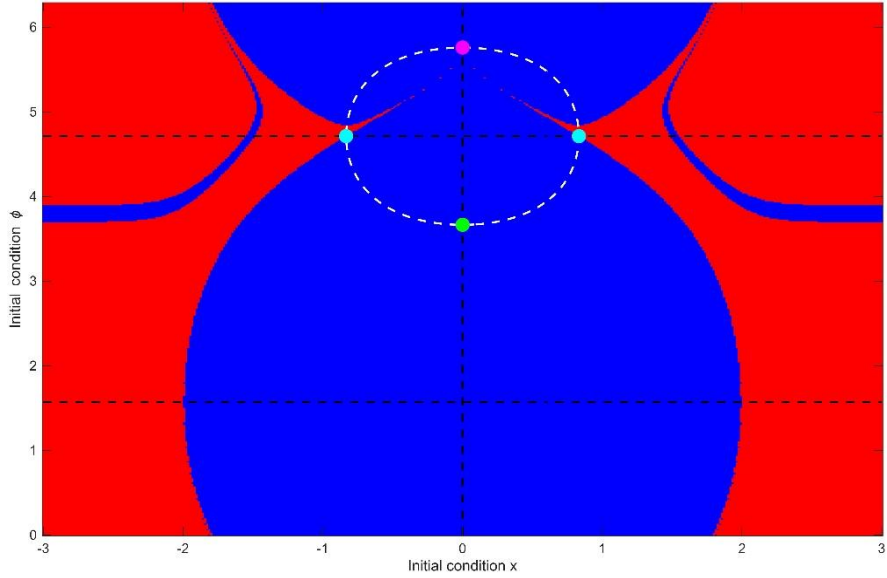

**Figure A2** Basin of attraction map for the Helical Excitable Dipole system for $\Delta=0.5$. The color overlay indicates the final long-term behavior based on initial conditions. The blue region denotes initial conditions that converge to the stable, stationary locked state (green circle). The red region indicates initial conditions that result in unbounded phase rolling. The dashed lines represent the nullclines, white for $\phi$, black for $x$. The fixed points are marked by circles: green for the stable node and magenta for the unstable node, cyan for saddle fixed points.

The introduction of a non-zero frequency detuning (termed anisotropy in the main text) $\Delta$ explicitly breaks the phase-reflection symmetry shown in Fig. A1. This perturbation induces a directed motion, effectively creating an Adler-like ratchet potential that is the precursor to the system's helical excitability. In order to further characterize the nature of the bifurcation from the locked to the unlocked state, I have numerically computed the oscillation period $\tau = 2\pi/\dot{\phi}$ as a function of the anisotropy parameter $\Delta$ for the HED. The results, presented in Figure A3, confirm the existence of a sharp transition around $\Delta_c = 1$. In the unlocked regime ($\Delta > 1$), $\tau$ follows a scaling law well described by the function

$$\tau = A/\sqrt{\Delta - 1} \qquad (A12)$$

This square-root dependence of the period on the distance from the critical point is a characteristic signature of a Saddle-Node bifurcation, which is a common mechanism for the onset of synchronization and oscillations in non-linear dynamical systems.

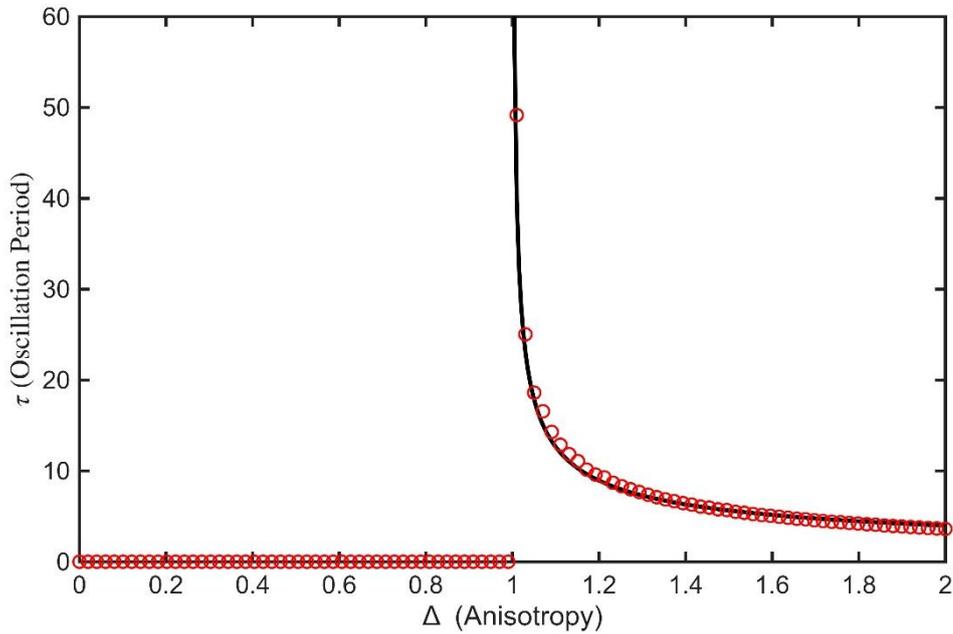

**Figure A3.** Oscillation period as a function $\tau$ of the anisotropy parameter $\Delta$ for the Helical Excitable Dipole. The red circles represent the data obtained from numerical simulations. The continuous black line is a fit curve following the scaling law Eq. (A12), highlighting the divergence of the period near the critical bifurcation point $\Delta_c = 1$.

As the parameter $\Delta$ approaches the threshold, the trajectory slows down in a "bottleneck" region. The time required to complete a cycle (the period) diverges to infinity following the inverse square-root law. In the context of biophysics and nonlinear dynamics, this corresponds to Type I excitability.

## Appendix B. Interdipolar Forces

Numerical simulations in Section 3.2 reported a regular pattern of almost equispaced HEDs, which diffuse spatially due to a repulsive interdipolar force (Type 1 solution, see Fig. 3). This appendix is dedicated to the analysis of this force. To evaluate it, I consider two HEDs separated by a distance D and I numerically calculate the radial component of the net force between them. This measures the effective interaction that drives their separation or approach. The simulation comprises N=4 COs (2 red and 2 blue) placed randomly within a small initial radius of 0.5 units from the origin, evolving according to Eqs. (9)-(10). The COs quickly form two dipoles that start to interact. I use a fixed random seed to ensure the reproducibility of these initial positions across all simulations for different $\Delta$ values, allowing for a consistent comparison of the resulting force profiles.

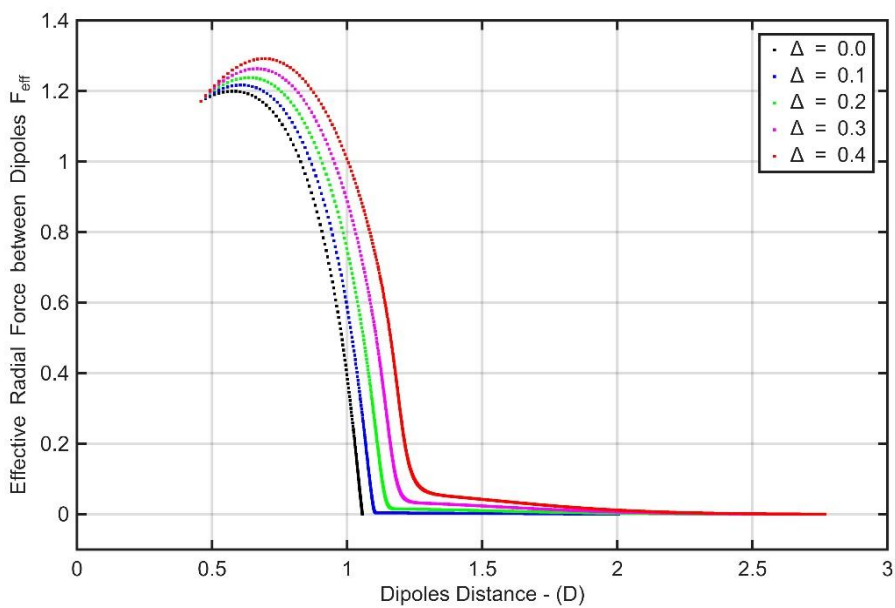

**Figure B1**. Effective radial force $F_{eff}$ as a function of the interdipolar distance D for different anisotropy parameters $\Delta$. For $\Delta = 0$, the force drops strictly to zero at larger distances, leading to ramified patterns in the many-body case. For non-zero $\Delta$ values, a weak residual repulsion persists, driving the crystalline arrangements observed in Figure 3 of the main text.

Figure B1 displays the effective radial force $F_{eff}$ as a function of the interdipolar distance D for various anisotropy parameters $\Delta$. The results demonstrate a distance-dependent interaction profile. Short Range Repulsion: The repulsive force increases significantly when the distance D is less than the characteristic interaction length scale L=1. Force Decay: As the distance increases beyond D ≈ 1 the repulsive force rapidly decreases. For $\Delta = 0$ (black dotted line), the force drops strictly to zero at larger distances. This total cessation of repulsion allows the formation of the ramified patterns, described in the main text (beginning section 3.2), as dipoles can stop interacting entirely when sufficiently separated. For non-zero $\Delta$ values (colored lines), the repulsive force persists, albeit weakly, even at larger distances. This residual repulsion prevents the indefinite aggregation seen at $\Delta = 0$ and drives the system towards the ordered crystalline arrangements reported in Figure 3 of the main text. While this pairwise approach neglects potential many-body effects, it provides a reliable qualitative description for sufficiently spaced dipoles, consistent with the dipolar interactions observed at low or vanishing anisotropy.